\documentclass[twocolumn,aps,prl,floatfix,superscriptaddress,nobibnotes,nofootinbib,altaffilsymbol]{revtex4-2}
\usepackage{amsmath,amssymb}
\usepackage{graphicx,float}
\usepackage{times,txfonts}
\usepackage{color}
\usepackage{multirow}
\usepackage{bbold}
\usepackage{color}
\usepackage{soul}
\usepackage{hyperref}
\usepackage{times,txfonts}
\usepackage{natbib}
\usepackage{nicefrac}
\usepackage{blindtext}
\usepackage[export]{adjustbox}
\usepackage{mathrsfs}
\usepackage{textcomp}
\usepackage{blkarray}
\usepackage{multirow}
\usepackage[normalem]{ulem}
\usepackage{tabu}

\renewcommand{\epsilon}{\varepsilon}

\def\VR{\kern-\arraycolsep\strut\vrule &\kern-\arraycolsep}
\def\vr{\kern-\arraycolsep & \kern-\arraycolsep}
\usepackage{sistyle}

\usepackage{soul}
\definecolor{lightblue}{RGB}{185,210,248}
\sethlcolor{lightblue}

\begin{document}

\title{Dynamic Topological Light Control in Reconfigurable Non-Hermitian Metastacks}

\author{Ryan Hogan$^\ddagger$}
\email{ryan.hogan@nju.edu.cn}
\affiliation{National Key Laboratory of Solid State Microstructures, School of Physics, and Collaborative Innovation Center of Advanced Microstructures, Nanjing University, Nanjing, Jiangsu, 210093, China.}

\author{Zihan Lu$^\ddagger$}
\affiliation{National Key Laboratory of Solid State Microstructures, School of Physics, and Collaborative Innovation Center of Advanced Microstructures, Nanjing University, Nanjing, Jiangsu, 210093, China.}

\author{Han Peng}
\affiliation{National Key Laboratory of Solid State Microstructures, School of Physics, and Collaborative Innovation Center of Advanced Microstructures, Nanjing University, Nanjing, Jiangsu, 210093, China.}

\author{Jie Zhou}
\affiliation{National Key Laboratory of Solid State Microstructures, School of Physics, and Collaborative Innovation Center of Advanced Microstructures, Nanjing University, Nanjing, Jiangsu, 210093, China.}

\author{Yanling Xiao}
\affiliation{National Key Laboratory of Solid State Microstructures, School of Physics, and Collaborative Innovation Center of Advanced Microstructures, Nanjing University, Nanjing, Jiangsu, 210093, China.}

\author{Shining Zhu}
\affiliation{National Key Laboratory of Solid State Microstructures, School of Physics, and Collaborative Innovation Center of Advanced Microstructures, Nanjing University, Nanjing, Jiangsu, 210093, China.}

\author{Hui Liu}
\email{hui.liu@nju.edu.cn}
\affiliation{National Key Laboratory of Solid State Microstructures, School of Physics, and Collaborative Innovation Center of Advanced Microstructures, Nanjing University, Nanjing, Jiangsu, 210093, China.}

\begin{abstract}
Metasurfaces often require complex lithography for dynamic optical control. To overcome this, we utilize a lithography-free, non-Hermitian planar metastack comprising a distributed Bragg reflector and a vanadium dioxide (VO$_2$) thin film. By virtue of temperature and thermal hysteresis as an active synthetic dimension and exploiting the VO$_2$ insulator-to-metal transition, we actively tune topological interface states to achieve polarization-sensitive spectral control. Notably, our system hosts path-dependent exceptional points (EPs); the intermediate hysteretic states generate a continuum of hot and cold EP pairs that ultimately converge into a single, degenerate EP. Furthermore, we experimentally observe wide-range dynamic optical control, comprising reversible 8\% spectral shifts with near-unity reflectance modulation, alongside potential for ultrafast dynamics. Ultimately, our CMOS-compatible design provides a scalable, simple platform for active and topological photonics.
\end{abstract}

\maketitle

\def\thefootnote{$\ddagger$}\footnotetext{These authors contributed equally to this work}\def\thefootnote{\arabic{footnote}}

\section{Introduction}
Over the past decade, metasurface (MS) design and implementation have advanced substantially \cite{ref01_0,ref01_1,ref01_2}, demonstrating vast functionality over many spectral regions \cite{ref02_1,ref02_2,ref02_3}. These designs compactify large, bulky optical systems into flat and ultrathin devices \cite{ref03_1,ref03_2,ref03_3}. However, as device footprints decrease and demand for higher integration grows, the reliability of fabrication poses a challenge \cite{ref04_1,ref04_2}. Particularly, high-precision optical control realized by active photonic devices succumbs to nanopatterning technical challenges and sensitivities \cite{ref05_1,ref05_2}. For high-level optical control in modern optics, the goal is simple for active photonics. That is to say, competitive alternatives such as CMOS-compatible, lithography-free platforms that replicate or surpass current functionality are necessary to advance the field, opening a new path for active photonic design structure and implementation with scalability as the design principle.

\begin{figure*}
    \centering
\includegraphics[width=\textwidth]{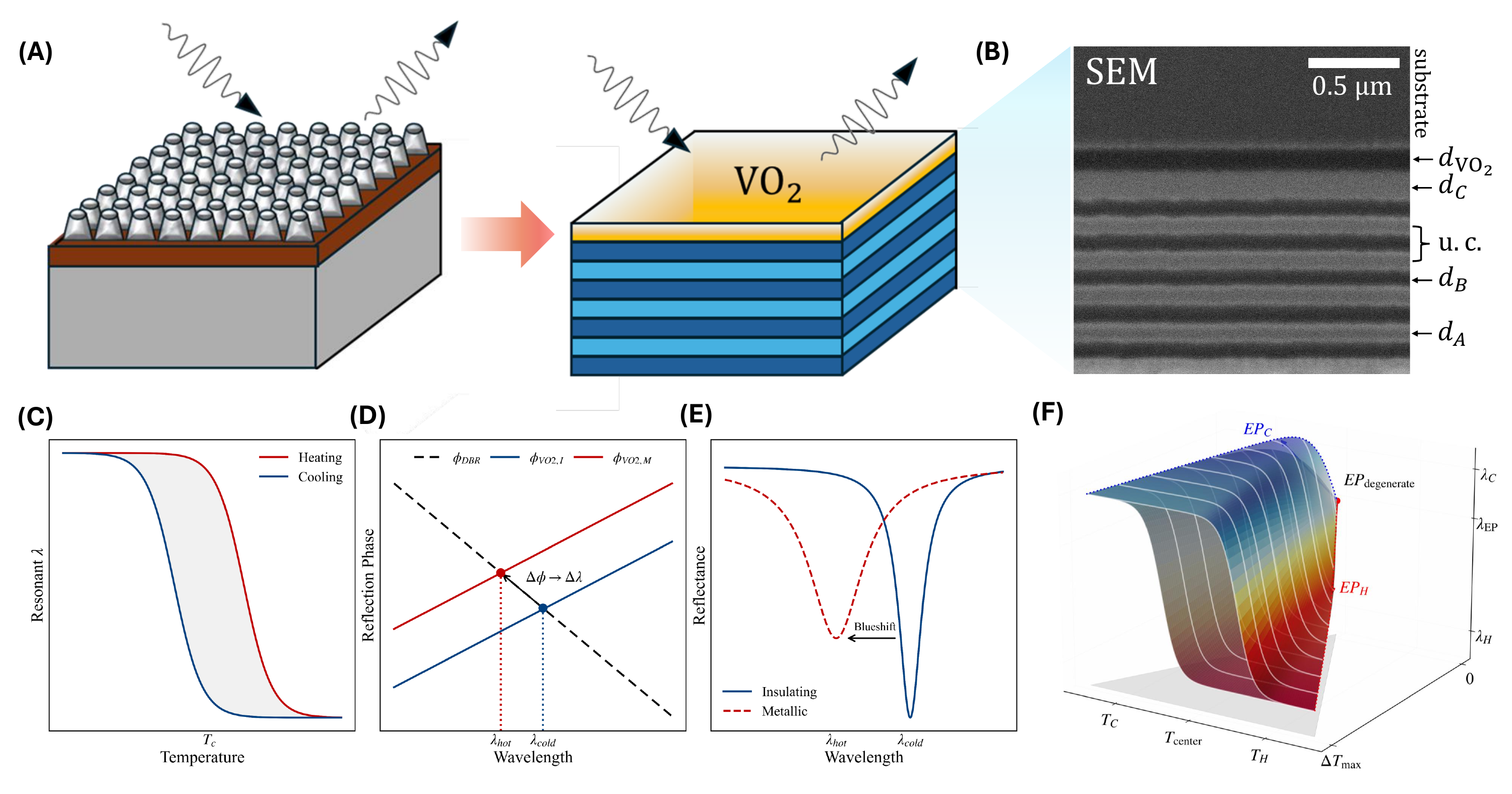}
    \caption{\textbf{Lithography-free metastack design and its projected functionalities.} 
\textbf{a.} Typical metasurface (MS) design (e.g., metallic nanopillars on a dielectric on glass), compared with our lithography-free $\mathrm{VO_2}$-Tamm metastack, consisting of thin-film $\mathrm{VO_2}$ on a distributed Bragg reflector (DBR) comprised of $\mathrm{Ta_2O_5}$ ($d_A$, $d_C=1.57d_A$) and $\mathrm{SiO_2}$ ($d_B$). 
\textbf{b.} Corresponding SEM highlighting layer thicknesses ($d_A, d_B$), unit cells (u.c.), the defect layer ($d_C$), the $\mathrm{VO_2}$ layer ($d_{\mathrm{VO_2}}$), and the substrate (quartz). 
\textbf{c.} Dramatic $\mathrm{VO_2}$ optical property changes strongly interact with optical Tamm states (OTSs) at the $\mathrm{VO_2}$/DBR interface, producing large resonant wavelength blue (red) shifts upon heating (cooling). 
\textbf{d.} Schematic of $\mathrm{VO_2}$ reflection phase response in insulating (metallic) states compared with the DBR, with OTS conditions met indicated in blue (red) dots. Corresponding change in phase is analogous to resonant wavelength shifts. 
\textbf{e.} Phase transition dynamics control OTS positioning in the stop band, leading to a large blue shift and spectral broadening. The $\mathrm{VO_2}$ plasma frequency blue shifts upon phase transitioning, leading to reflection phase condition dynamic control and impedance mismatch control of constituent layers. 
\textbf{f.} 3-D topological landscape showing how path-dependent exceptional points ($\text{EP}_H$ and $\text{EP}_C$) converge and merge into a single $\text{EP}_{\text{degenerate}}$ as the hysteresis width ($\Delta T_{\text{max}} \rightarrow 0$) collapses.}
    \label{fig:Figure1}
\end{figure*}

In general, MS dynamic control is realized through reconfigurable devices, utilizing platforms such as graphene \cite{ref06_Graphene} or active materials (i.e., GST, VO$_2$) \cite{ref07_1_GST,ref07_2_GST,ref08_1_VO2,ref08_2_VO2}. Yet, high-resolution lithography is often necessary, remaining both specialized and cost-intensive, significantly impacting the overall quality of the optical response \cite{ref09_1,ref09_2}. As a result, we turn to resonances that occur in a planar stack design approach. Optical Tamm States (OTSs) are resonant features offering a compelling alternative to fabrication-intensive MSs. OTS resonant interface modes are well-studied and useful in diverse applications \cite{ref10_1,ref10_2,ref10_3,ref10_4}. The topological properties of OTS are investigated with Zak phase measurement and synthetic dimension \cite{ref12_2,wang2017optical}. However, OTSs are typically limited to static response dictated by the fixed metal-dielectric boundary \cite{ref11_1,ref11_2,ref11_3}. By substituting the metal layer with a thin film of vanadium dioxide (VO$_2$), we create a dynamic, reconfigurable environment from which we can optically control the OTSs. This approach serves as a functional analog to nanostructured MSs where spatial fill-fractions are used to control resonance and $Q$-factor \cite{ref12_1}; however, while these properties are permanently fixed by fabrication geometry, the phase co-existence during the VO$_2$ insulator-to-metal transition (IMT) allows for continuous optical control and spectral tuning. Moreover, as VO$_2$ exhibits thermal hysteresis, the dynamics on heating and cooling differ, and therefore allow for a wider range of tunability. On heating, as the metallic phase grows, OTS resonance shifts according to the VO$_2$ plasma frequency shift. Moreover, the difference in paths allows for distinct topological physics to exist, specifically the existence of exceptional points on heating $\text{EP}_H$ and on cooling, $\text{EP}_C$. When examining temperature and thermal hysteresis as a synthetic dimension, a plethora of new topological physics can be observed. By tuning through the IMT, the critical coupling condition can be achieved, and coupling can be tuned in real time, which collapses the EP degeneracy splitting.

Unlike phase-change materials (PCMs) such as GST, which are limited to discrete switching \cite{ref13_1,ref13_2}, VO$_2$ allows for predictable, continuous, energy-efficient control near room temperature. Moreover, owing to the thermal, electrical, or optical excitation methods, VO$_2$ is quite versatile and has potential for ultrafast modulation on sub-picosecond timescales \cite{ref14_1,ref14_2,ref14_3}. Overall, the structure is simple and lithography-free while also achieving high-contrast tunability. Hereafter, we denote our structure as a metastack, which signifies its many-sided functionality akin to metasurfaces, and its structural form as a planar stack.

In this work, we demonstrate the vast performance of the VO$_2$-driven OTS metastack, here designed near 800 nm. We achieve a reversible spectral shift of approximately 60 nm across the VO$_2$ IMT phase transition, consistent with transfer-matrix and analytical models. This broad spectral response is accompanied by a near-unity modulation efficiency of 82\% in reflection. Moreover, we present the platform’s polarization response and dispersive effects as further degrees of tunability. Alongside the dispersion, we also show that, although OTSs need not exhibit topological properties as a result of the bandgap of the constituent materials, the VO$_2$-driven OTS does, leading to polarization-sensitive EPs. Introducing temperature as a synthetic dimension, we show that exceptional points can be designed for specific angles and temperatures, yielding distinct topologies, thereby enabling wide tunability of the system’s non-Hermitian, topological properties.

Overall, our design offers a compact, CMOS-compatible structure that supports a new pathway for active photonic devices. Our design shows promise for future application in ultrafast modulation limited only by the VO$_2$ electronic lifetime. The scalability and planar nature of the metastack are excellent for adaptive optics, flat-optical devices, and integrated devices \cite{ref15_1,ref15_3,ref15_4}, showing potential for high-speed switching \cite{ref16_1,ref16_2,ref16_4}, beam steering \cite{ref17_1,ref17_2}, dynamic optical filtering \cite{ref18_1,ref18_2}, and optical communications \cite{ref19_1}. Moreover, the underlying physics is inherently broadband, suggesting a clear path toward MIR \cite{ref20_1,ref20_2} and THz functionalities \cite{ref21_1,ref21_2}.

\section{Results}
We begin by introducing the concept of dynamic control of our metastack as an alternative to MSs (Fig. \ref{fig:Figure1}\textbf{a}). While typical MSs rely on the geometric arrangement of nanostructures to achieve optical control, our structure reaches competitive functionality through the intrinsic phase-change dynamics of a $\mathrm{VO_2}$ thin film. Our corresponding fabricated device (Fig. \ref{fig:Figure1}\textbf{b}) consists of a 63-nm $\mathrm{VO_2}$ layer placed at the interface of a $\mathrm{Ta_2O_5}$/$\mathrm{SiO_2}$ distributed Bragg reflector (DBR). More specifically, a $\mathrm{VO_2}$ layer is fabricated on top of quartz glass, upon which a DBR is constructed. Our specific design targets a central wavelength of $\lambda_c = 800$ nm, utilizing symmetric DBR unit cells $[d_A/2, d_B, d_A/2]$ repeated over four periods, where $d_{A,B}=\lambda_c/(4n_{A,B})$. The sequence terminates with an additional defect cell, $[d_A/2, d_B, d_C]$, where $d_C = 1.57 d_A$ denotes the defect-layer thickness, optimized to balance resonance depth with practical fabrication tolerances. 

\begin{figure*}
    \centering
    \includegraphics[width=\textwidth]{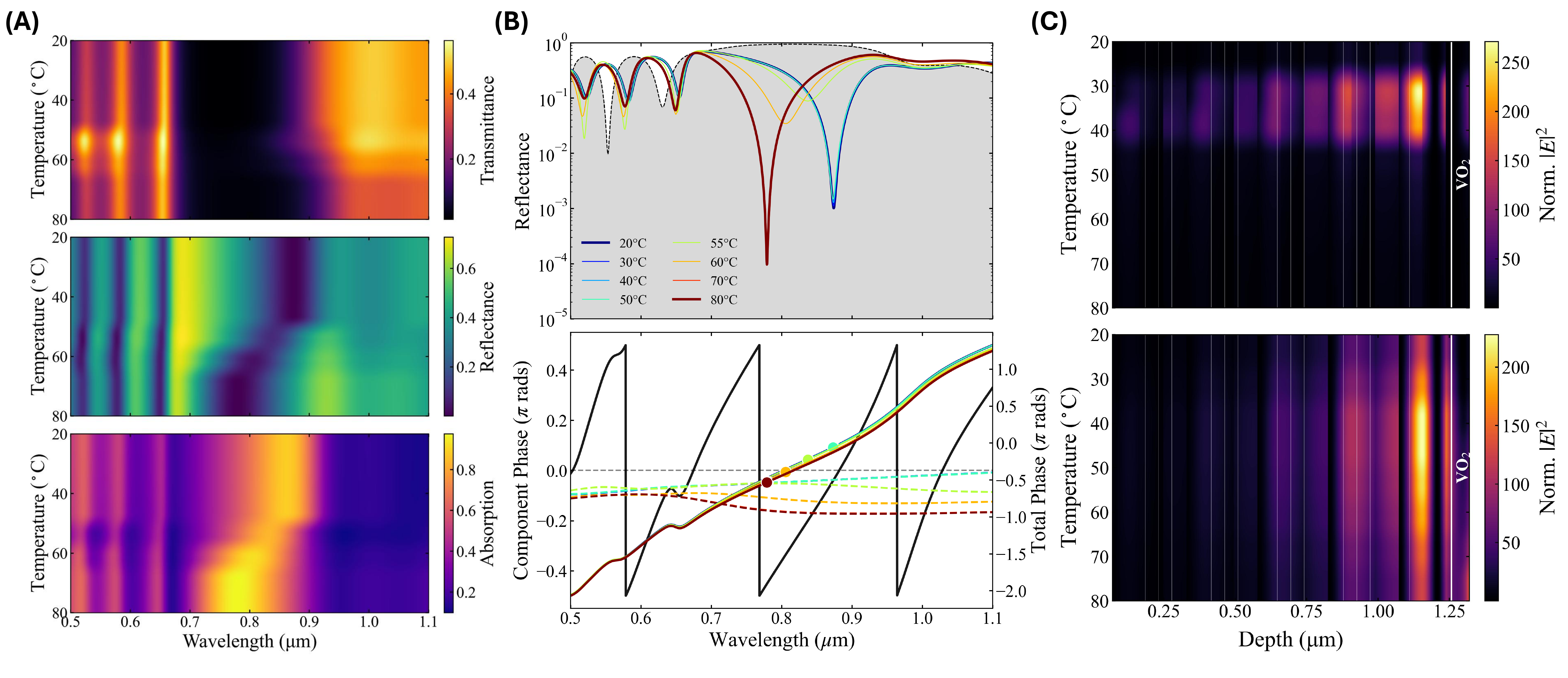}
    \caption{\textbf{Numerical characterization of $\mathrm{VO_2}$-OTS metastack.} 
\textbf{a.} Simulated transmittance (top), reflectance (middle), and absorption (bottom) spectral response map temperature and wavelength dependence for our metastack. The OTS shifts from $\lambda_{\mathrm{OTS},C} \approx 880$ nm to $\lambda_{\mathrm{OTS},H} \approx 780$ nm, giving $\Delta\lambda \approx 100$ nm. Light is well localized, reaching $\sim 90\%$ absorption. 
\textbf{b.} Temperature-dependent reflectance is plotted, highlighting OTS linewidths (DBR in grey). The defect layer optimized OTS position and depth, $d_C$, however, fabrication tolerances are $5\%$ of desired thicknesses for deposition of layers necessitating $\lambda/8$. Corresponding component reflection phase (scaled by $1/2$) is shown for the DBR (in black) and $\mathrm{VO_2}$ only (in dashed colored lines) on the left axis, and their sum on the right axis (solid colored lines). Colored dots denote OTS reflectance dip positions. 
\textbf{c.} Field evolution (left illuminated on DBR facet) is monitored, showing strong confinement at hot and room temperature for $\lambda_{\mathrm{OTS},C} \approx 780$ nm (localized for all temperatures), and $\lambda_{\mathrm{OTS},H} \approx 880$ nm (localized for low temperatures due to small $k$).}
    \label{fig:Figure2}
\end{figure*}

Schematically, the expected dynamics of the metastack are depicted in Figs. \ref{fig:Figure1}\textbf{c}-\textbf{e}. Figure \ref{fig:Figure1}\textbf{c} shows shifting of the resonant wavelength as a function of temperature, albeit also capable of control via voltage, optical pulses, among others. As $\mathrm{VO_2}$ exhibits thermal hysteresis, the dynamics of the shifting follow suit. By driving the $\mathrm{VO_2}$ through its IMT, we fundamentally reconfigure the photonic boundary conditions at the $\mathrm{VO_2}$-DBR interface, thereby dynamically tuning the OTS resonant wavelength. The modified reflection phase boundary conditions are schematically depicted, showing the phase-matched conditions (Fig. \ref{fig:Figure1}\textbf{d}). As the metallic phase grows, the resulting blue shift of the plasma frequency enables real-time control of the resonance position. Hereby, the spectral position and line shape are modified as a result of the modified impedance matching (Fig. \ref{fig:Figure1}\textbf{e}). Consequently, we can then identify critical coupling conditions required for near-unity modulation. 

Figure \ref{fig:Figure1}\textbf{f} illustrates the resulting non-Hermitian topological space mapped across the synthetic dimension temperature ($T$), hysteresis width ($\Delta T$), and resonance wavelength ($\lambda$). When operating along the complete hysteresis loop ($\Delta T_{\text{max}}$), the system hosts two distinct, path-dependent topological singularities, $\text{EP}_H$ along the heating trajectory (red) and $\text{EP}_C$ along the cooling trajectory (blue). As we drive incomplete thermal transitions to collapse the hysteresis width ($\Delta T \rightarrow 0$), these path-dependent paths continuously contract, forcing two isolated EPs to converge into a single, path-independent $\text{EP}_{\text{degenerate}}$ node at the tip of the synthetic parameter space. Next, a rigorous theoretical framework for the physics of the phase-driven resonance shift is established using a transfer-matrix method (TMM) approach, followed by temperature-dependent reflectance measurements of the fabricated device for experimental validation, revealing high-contrast, reversible spectral tunability and modulation.

To model our metastack system, we first establish the conditions for an OTS at the $\mathrm{VO_2}$-DBR interface. That is,
\begin{equation}
    r_{\mathrm{DBR}} \cdot r_{\mathrm{VO_2}} \cdot e^{2i\phi_{tot}} = 1,
    \label{eq:Eq1}
\end{equation}
In general, the OTS emerges when the round-trip phase completes a $2\pi$ condition, as described by
\begin{equation}
    \phi_{tot} = \phi_{\mathrm{DBR}} + \phi_{\mathrm{VO_2}}= 2 \pi m, \quad m \in \mathbb{Z}.
    \label{eq:Eq2}
\end{equation}

Coupled to the $\mathrm{VO_2}$ electronic state change during IMT, Eq.~\eqref{eq:Eq2} conditions shift continuously with the plasma frequency shift, wherein different $\mathrm{VO_2}$ reflection phases exist. The optical impedance evolves in tandem. A benefit of continuous reconfiguration is the ability to tune boundary conditions to achieve critical coupling, where the optical field becomes strongly confined.

\begin{figure*}
    \centering   
    \includegraphics[width=\textwidth]{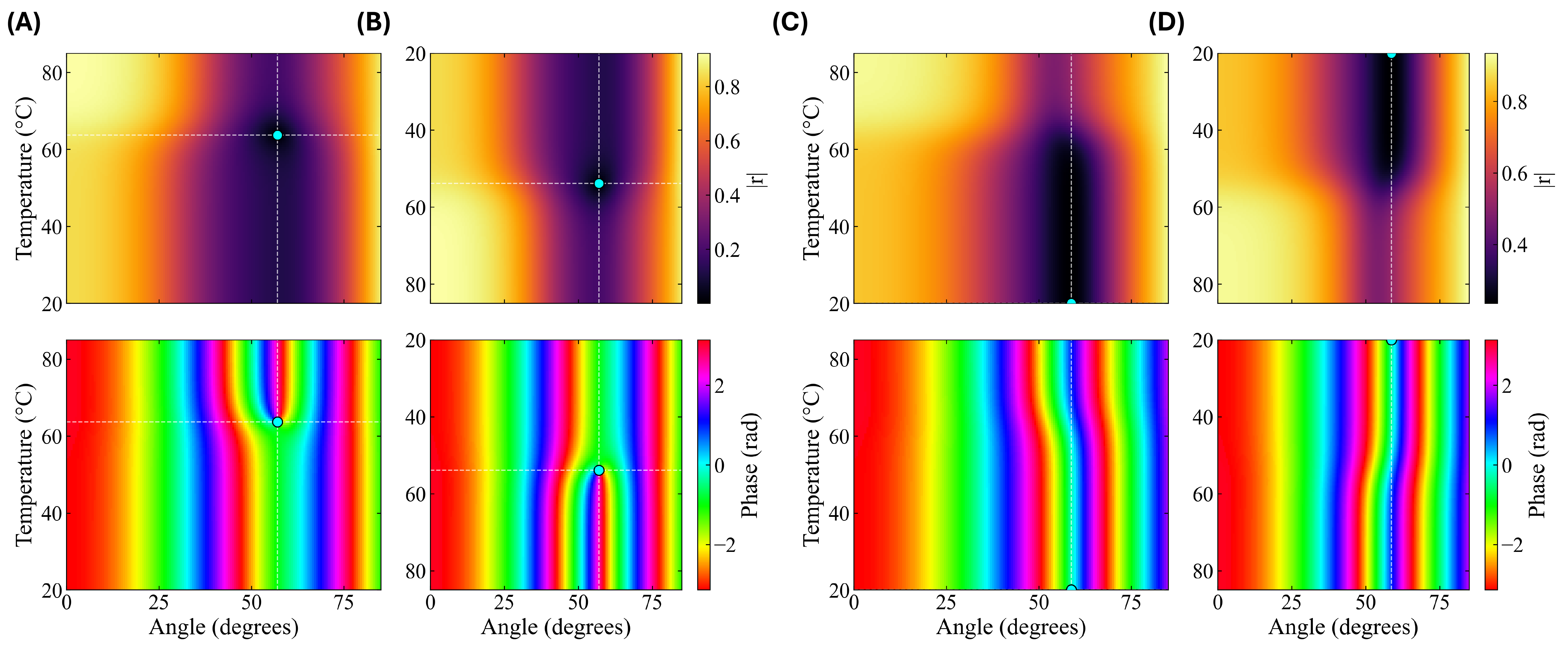}
    \caption{\textbf{$\mathrm{VO_2}$-OTS Exceptional Points and Polarization Response.} 
The metastack reflection coefficient magnitude (top) and phase (bottom) are plotted \textbf{a.} for heating and \textbf{b.} for cooling. The target wavelength for the simulated behavior is set at the resonance position of the OTS in the dielectric phase of the $\mathrm{VO_2}$ layer, corresponding to $\lambda_{\mathrm{OTS},d} \approx 880$ nm. The minimum reflection magnitude is denoted with a cyan dot. Here, the reflection amplitude approaches zero, which is reflected in the phase vortex behavior seen in the reflection phase. At a temperature of $T_{EP} = 63.7^\circ$C and an angle of $\theta_{EP} = 57.5^\circ$ marks $\text{EP}_H$ in the temperature-angle synthetic space. The corresponding $\text{EP}_C$ is also highlighted in cyan, at a lower temperature near $T_{EP,C} \approx 57.5^\circ$C. 
\textbf{c.} The corresponding s-polarization reflection coefficient magnitude and phase are plotted for heating, showing that an exceptional point is no longer observed, indicating polarization-sensitivity. The cyan dot marks the minimum amplitude, attaining a value of $|r| \sim 0.3$. 
\textbf{d.} Similar results for s-polarization on cooling are shown.}
    \label{fig:Figure3}
\end{figure*}

To capture this active control, we model the $\mathrm{VO_2}$ permittivity as a temperature-dependent interpolation of its dielectric and metallic states. We employ a sigmoidal weighting function, $\rho(T)$, to describe the evolution of the dielectric permittivity of the thin film $\mathrm{VO_2}$, $\varepsilon_{\mathrm{VO_2}}$, during IMT:
\begin{equation}
    \varepsilon_{\mathrm{VO_2}}(T, \lambda) = \rho(T)\epsilon_{\mathrm{H}}(\lambda) + [1 - \rho(T)]\epsilon_{\mathrm{C}}(\lambda),
    \label{eq:Eq3}
\end{equation}
where ``C'' (``H'') denotes the dielectric (metallic) states (otherwise noted as cold and hot states), $\lambda$ is the free-space wavelength, and $\rho(T) = \left(1+e^{-\alpha (T - T_c)}\right)^{-1}$, with $\alpha$ as a scaling factor, and $T_c$, the phase transition temperature. This function accounts for the typical smooth hysteresis of thin film $\mathrm{VO_2}$ \cite{ref08_1_VO2,ref08_2_VO2}.

While the full optical response is calculated via TMM, the resonance wavelength ($\lambda_{\mathrm{Tamm}}$) can be intuitively extracted from the interface phase-matching condition by numerically solving Eq.~\eqref{eq:Eq2}. This approach confirms that the constructive interference required to confine the OTS is maintained throughout the entire reconfiguration process.

The phase contribution from the $\mathrm{VO_2}$ layer, $\phi_{\mathrm{VO_2}}$, is directly governed by the impedance mismatch at the interface:
\begin{equation}
    \phi_{\mathrm{VO_2}} = \arg\left( \frac{\sqrt{\epsilon_{\mathrm{inc}}(\lambda)} - \sqrt{\epsilon_{\mathrm{VO_2}}(T, \lambda)}}{\sqrt{\epsilon_{\mathrm{inc}}(\lambda)} + \sqrt{\epsilon_{\mathrm{VO_2}}(T, \lambda)}} \right),
    \label{eq:Eq4}
\end{equation}
where $\epsilon_{\mathrm{inc}}$ is the permittivity of the adjacent DBR layer. This framework ensures that the constructive interference required to confine the OTS is maintained throughout the transition, providing a predictive model for the continuous spectral tuning we observe. Such a model directly links the material’s electronic state to its optical response. By quantifying the high-contrast tuning of the optical coupling strength at the interface, we identify the fundamental mechanism behind the near-unity modulation realized in our experiments.

\begin{figure}[t!]
    \centering
\includegraphics[width=\columnwidth]{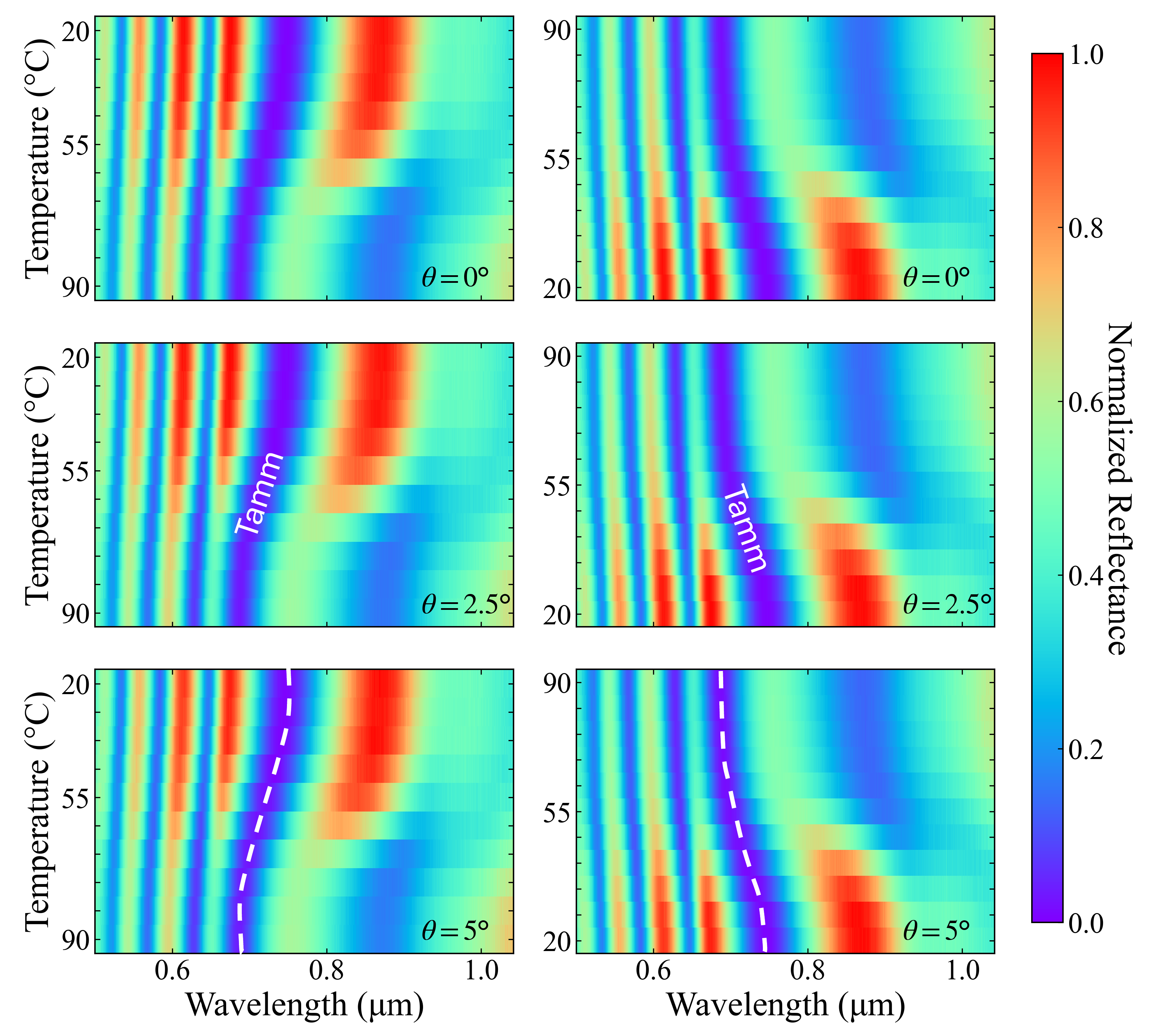}
    \caption{\textbf{Active spectral tunability of $\mathrm{VO_2}$-OTS resonances upon heating and cooling.} The spectral response of reflectance is measured at ten different temperatures, heating from $T=[20, 30, 40, 50, 55, 60, 65, 70, 80, 90]^\circ$C, for three incident angles, $\theta_i = [0^\circ, 2.5^\circ, 5^\circ]$. A shift of the OTS is observed from $\lambda_{\mathrm{OTS},C} = 755$ nm, to $\lambda_{\mathrm{OTS},H} = 695$ nm, where $C$ and $H$ refer to $\mathrm{VO_2}$ cold (dielectric) and hot (metallic) states within the stack, resulting in a $\Delta\lambda \approx 60$ nm. Likewise, eleven temperatures are measured upon cooling, from $T=[90, 80, 70, 65, 60, 55, 50, 45, 40, 30, 20]^\circ$C, and curves shift from $\lambda_{\mathrm{OTS},H} = 695$ nm, to $\lambda_{\mathrm{OTS},C} = 755$ nm, indicating the continuous and reversible control of the shift. Note that the experimental data shift is slightly smaller than numerical calculations due to the use of online data (More details on numerical analysis using our $\mathrm{VO_2}$ thin films can be found in Supplementary Materials). Note that the shift dynamics are different on heating and on cooling, indicative of $\mathrm{VO_2}$ thermal hysteresis. The simulated response of the angular dispersion can be found in the Supplementary (Fig. S3).}
    \label{fig:Figure4}
\end{figure}

While the resonant wavelength can be extracted directly from the interface phase-matching condition, we employ the full TMM formalism to capture the full spectral response and line-shape evolution. This approach accounts for the finite thickness of the $\mathrm{VO_2}$ layer and the wavelength-dependent losses across the transition. The resulting numerical simulations in Fig. \ref{fig:Figure2} fully characterize the optical response, providing a predictive foundation for the experimental findings discussed later.

As shown in the spectral maps in Fig. \ref{fig:Figure2}\textbf{a}, our metastack geometry yields a broad photonic stopband spanning 680 nm to 920 nm (shown in grey at the top of Fig. \ref{fig:Figure2}\textbf{b}). More specifically, the reflectance shows a characteristic dip within the stopband, indicating the dynamic evolution of the OTS. At room temperature, the resonance is located at $\lambda_{\mathrm{OTS}, C} \approx 880$ nm, but upon heating, it undergoes a dramatic blueshift of $\Delta\lambda \approx 100$ nm, settling at $\lambda_{\mathrm{OTS}, H} \approx 780$ nm (Fig. \ref{fig:Figure2}\textbf{b}), all within the stopband. Throughout this transition, the mode maintains high interaction efficiency with absorption rates reaching $\sim 90\%$ at the resonance positions (bottom of Fig. \ref{fig:Figure2}\textbf{a}). The corresponding reflection phase of the components is plotted on the left axis for the DBR (black), and $\mathrm{VO_2}$ only (colored, dashed lines) for different temperatures. Note that the component phases are scaled by half. The total reflection phase (i.e., the sum of the constituents in colored solid lines) is plotted on the right axis. Colored dots indicate positions of the reflectance dips associated with the OTS for each temperature. Finally, the spatial field distributions in Fig. \ref{fig:Figure2}\textbf{c} confirm the localized nature of the state. Whether in the insulating or metallic phase, the electric field exhibits strong enhancement within the terminal $\mathrm{VO_2}$ layer, displaying the characteristic exponential decay into the DBR expected of a bound Tamm mode.

Figure \ref{fig:Figure3} shows the reflection coefficient amplitude and reflection phases, for all four combinations of p- and s-polarizations, and upon heating and cooling. By operating at $\lambda_{\mathrm{OTS},d} \sim$ 880 nm, the resonance wavelength of the OTS when $\mathrm{VO_2}$ is in its fully dielectric phase, we can exploit this thermal tuning as a synthetic variable that creates a space capable of expressing exceptional points (EPs). Specifically, in the two-dimensional synthetic parameter space ($T,\theta$), the complex reflection coefficient is mapped, revealing a distinct topological singularity for p-polarized light in Fig. \ref{fig:Figure3}\textbf{a} upon heating. At the critical coordinates of $T_{EP} = 63.7^\circ$C and $\theta_{EP} = 57.5^\circ$C, the reflection amplitude (top) undergoes critical coupling and vanishes ($|r| \to 0$). This point is suggestive of a non-Hermitian spectral degeneracy associated with near-perfect absorption. Topologically, this singularity is accentuated by the abrupt phase vortex observed in the reflection phase map (Fig. \ref{fig:Figure3}\textbf{a}, bottom), confirming that the system has encircled an EP, here being the $\text{EP}_H$ discussed in Fig. \ref{fig:Figure1}\textbf{b}. The hysteretic nature of the $\mathrm{VO_2}$ transition dictates that the pathway to this topological feature depends strongly on the thermal history of the device, yielding distinct profiles for the heating and cooling branches. The corresponding p-polarized EP on cooling is shown in Fig. \ref{fig:Figure3}\textbf{b}, at a shifted set of coordinates in the $T-\theta$ space. Specifically, the existence of $\text{EP}_H$ and $\text{EP}_C$ strongly suggests an opening of a degeneracy for a non-zero thermal hysteresis width. That is to say, the area enclosed by thermal hysteresis in the synthetic $T-\theta$ space is capable of hosting a significant number of EPs. Moreover, expanding the space into a volumetric synthetic space, specifically $T,\lambda,\theta$-space, further grows the possible EP points. In Fig. \ref{fig:Figure3}\textbf{c} and \textbf{d}, which highlight s-polarization, EPs disappear. Reflection amplitude only reaches a shallow minimum of $|r| \sim 0.3$. The requirements to satisfy critical coupling conditions are disrupted such that the s-polarized OTS cannot simultaneously match the required radiation leakage and inherent material loss. As a result, we observe a highly polarization-sensitive EP behaviour, further adding depth of optical control. 

Following the numerical analysis, the tunability of the fabricated $\mathrm{VO_2}$-Tamm structure was experimentally characterized by monitoring the reflectance spectra across a complete thermal cycle. As illustrated in Fig. \ref{fig:Figure4}, measurements were recorded at three incidence angles ($\theta_i = 0^\circ, 2.5^\circ, 5^\circ$) while heating the sample from $20^\circ$C to $90^\circ$C. As the $\mathrm{VO_2}$ layer undergoes its IMT, the OTS resonance exhibits a clear, pronounced blueshift. For the $0^\circ$ case, the resonance moves from $\lambda_{\mathrm{OTS}, C} \approx 755$ nm to $\lambda_{\mathrm{OTS}, H} \approx 695$ nm, producing a significant spectral tuning range of $\Delta \lambda \approx 60$ nm. The corresponding positions of the OTSs within the DBR are shown in more detail in the Supplementary (Fig. S3).

To verify reversibility, the spectral response was subsequently tracked during cooling from $90^\circ$C to $20^\circ$C. The resonance continuously returned to its initial position, confirming the robust and reversible control of the Tamm state. Notably, the dynamics of the spectral shift differ between the heating and cooling phases, clearly revealing the thermal hysteresis loop characteristic of the $\mathrm{VO_2}$ phase transition. The specific resonance positions vary slightly with the angle of incidence, as expected by the simulated quadratic angular dispersion results (Fig. S2). While these measurements highlight the angle-sensitive nature of the device, a comprehensive experimental analysis of the full angular spectrum and polarization response remains a subject for future investigation. In general, our experimental results show good agreement with the predicted theoretical trends; however, simulations were performed with previous $\mathrm{VO_2}$ n/k literature values \cite{oguntoye2023continuously}, which predict slightly larger shifts of 100 nm in the OTS wavelength. Measured n/k values for our fabricated thin films can be found in Supplementary Materials (Fig. S1).

\begin{figure}[t!]
    \centering
\includegraphics[width=\columnwidth]{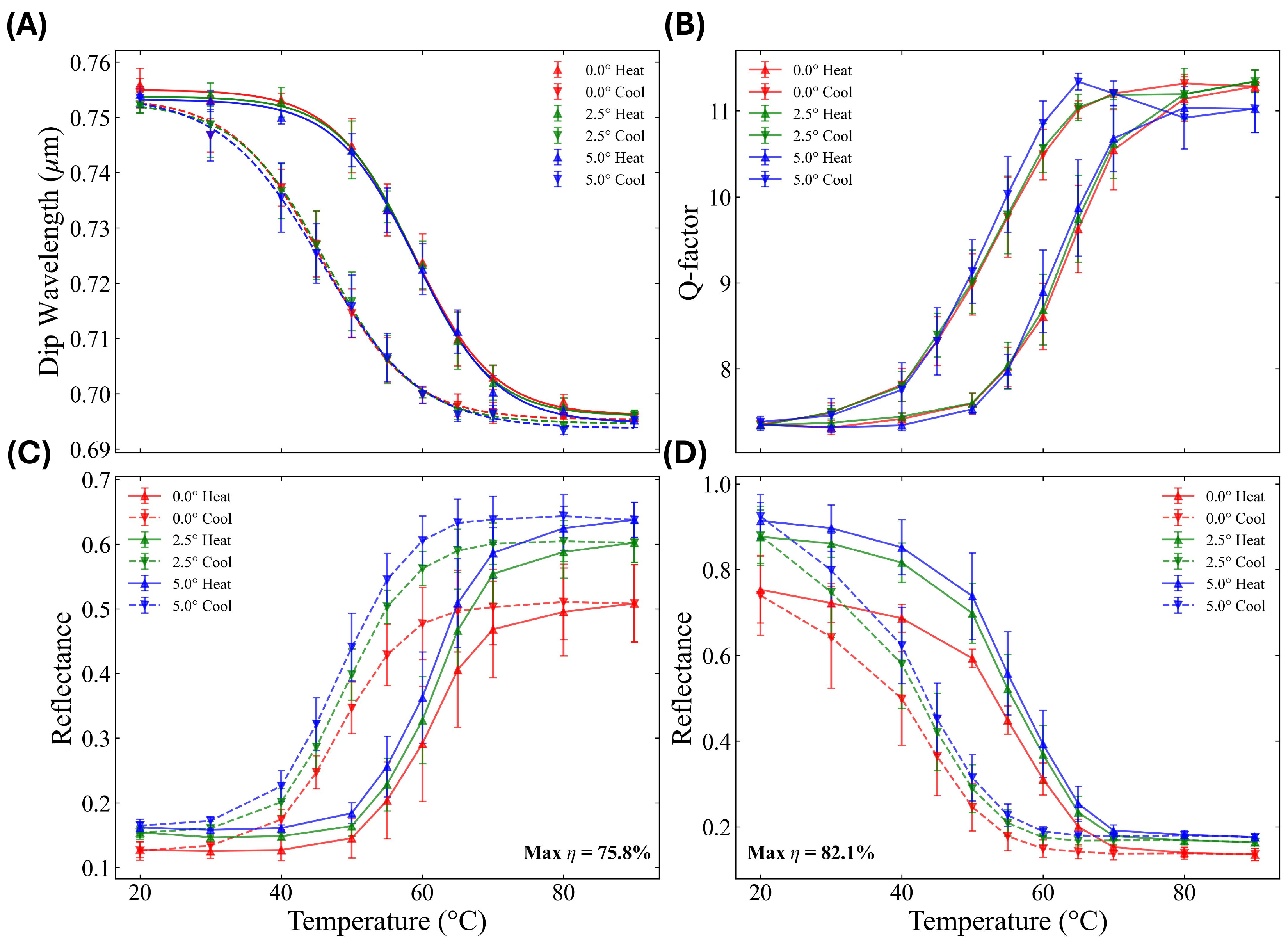}
    \caption{\textbf{High-contrast switching and thermal hysteresis.} \textbf{a., b.} Temperature dependence of the OTS resonant wavelength and Q-factor of the resonance, respectively, for heating (triangles) and cooling (inverted triangles). The sigmoidal fits (solid and dashed lines) reveal a phase transition temperature of $T_{cr} = 58.8\pm0.2^\circ$C. The average thermal hysteresis width $\Delta T_{hyst} = 12.4\pm0.4^\circ$C over measured angles is due to natural OTS dispersion. \textbf{c., d.} Modulation capabilities of the dielectric and metallic $\mathrm{VO_2}$-OTS modes, respectively. The dielectric mode achieves a maximum efficiency of $\eta_{C, \mathrm{max}} = 77.4\%$ ($R_{\mathrm{max}}=0.574, R_{\mathrm{min}}=0.130$). The metallic mode exhibits an enhanced modulation efficiency of $\eta_{H, \mathrm{max}} = 82.1\%$, with absolute reflectance extrema of $R_{\mathrm{max}}=0.771$ and $R_{\mathrm{min}}=0.142$. The robust field localization at these resonance positions is confirmed by the corresponding field maps in Fig. \ref{fig:Figure2}.}
    \label{fig:Figure5}
\end{figure}

To quantify the switching dynamics, the resonant wavelength (Fig. \ref{fig:Figure5}\textbf{a}) and quality factor (Fig. \ref{fig:Figure5}\textbf{b}) of the OTS were extracted from the reflectance spectra and plotted as a function of temperature for both heating and cooling cycles. The phase transition characteristics, including the critical temperature and hysteresis width, were derived by fitting the wavelength shift to a sigmoidal model. This analysis yielded a mean critical heating transition temperature of $T_{cr} = 58.8 \pm 0.2^\circ$C. By determining the corresponding cooling transition temperatures for each incidence angle, the average thermal hysteresis width was calculated to be $\Delta T_{hyst} = 12.4 \pm 0.4^\circ$C. The slight broadening of the hysteresis loop observed with increasing angle is attributed to the natural angular dispersion of the OTS mode rather than intrinsic variations in the $\mathrm{VO_2}$ optical properties.

The optical switching performance was further evaluated by monitoring the reflectance contrast at the specific resonant wavelengths associated with the dielectric ($\lambda_{\mathrm{OTS},C}$) and metallic ($\lambda_{\mathrm{OTS},H}$) states (Figs. \ref{fig:Figure5}\textbf{c} and \textbf{d}). For the dielectric $\mathrm{VO_2}$-OTS mode, the structure achieves a maximum efficiency of $\eta_{C, \mathrm{max}} = 77.4\%$, corresponding to a reflectance change from $R_{\mathrm{max}} = 0.574$ to $R_{\mathrm{min}} = 0.130$. Conversely, the metallic $\mathrm{VO_2}$-OTS mode demonstrates superior performance with an efficiency of $\eta_{H, \mathrm{max}} = 81.6\%$, driven by a dynamic range between $R_{\mathrm{max}} = 0.771$ and $R_{\mathrm{min}} = 0.142$. The robust field localization at these resonance positions, which underpins this high-contrast modulation, is confirmed by the corresponding field maps in Fig. \ref{fig:Figure2}.

\section{Conclusion}
In summary, we have demonstrated a lithography-free, planar stack that enables near-arbitrary active optical functionality utilizing a commonly known phase-change material, VO$_2$, in place of the metal at the interface with a DBR comprising our metastack. Through this platform, we realized a structure capable of supporting OTSs. These OTSs are highly sensitive to boundary conditions and can be thermally triggered to induce the IMT of VO$_2$, achieving a large spectral tuning range of 60 nm within the photonic stopband. The device exhibits robust optical switching, with a maximum modulation efficiency of 82.1\%. Moreover, we observed that by using temperature as a synthetic parameter, our platform exhibits topological exceptional points in synthetic space that are polarization sensitive. These results confirm that exploiting the interface phase-matching conditions of resonant modes offers a viable, low-cost alternative to complex nanostructured MSs.

Our VO$_2$-planar stack establishes a high-performance platform for active resonance control in VIS-NIR. The underlying physics has yet to be explored in depth, both fundamentally and through applied research. For instance, exploiting the subpicosecond IMT of VO$_2$  could transition this device from a thermal modulator into an ultrafast all-optical switch for high-speed data processing. Furthermore, a more detailed mapping and experimental characterization of the dispersion and polarization response characteristic of OTS modes may lead to highly sensitive angle-selective filters and sensors. Finally, the interaction between the lossy metallic state of VO$_2$  and the confined Tamm mode offers a unique environment for studying non-Hermitian photonics and exceptional points, providing a fundamentally new pathway toward high-speed, active photonic devices and their subsequent integration.\\
\vspace{3mm}

\noindent \textbf{Funding}\\
The authors acknowledge support from the National Natural Science Foundation of China (Grant Nos. W2533027, 12334015), and the Jiangsu Province Excellent Postdoctoral Fellowship Program (Grant No. 2025ZB498).
\vspace{3mm}

\noindent \textbf{Acknowledgements}\\
The authors would like to acknowledge Xiaoqin Gao, Ruoyang Zhang, and Kai Gao, and Jinguang Shang for valuable scientific discussions. 
\vspace{3mm}

\noindent \textbf{Author Contributions}\\
Conceptualization: RH\\
Methodology: RH, ZL, HP \\
Investigation: RH, ZL, HP, JZ\\
Fabrication and Characterization: JZ, ZL, RH, YX\\
Visualization: RH, ZL, HP\\
Formal Analysis: RH, ZL: RH, ZL\\
Funding acquisition: RH, HL\\
Supervision: HL, SZ\\
Writing – original draft: RH\\
Writing – review \& editing: All authors contributed.\\
\vspace{3mm}

\noindent \textbf{Competing Interests Statement}\\
The authors declare no competing interests.

\newpage 
\section{References}
\bibliography{Main/references}

\newpage

\section{Supplementary Materials}

The spectral response of the multilayer structure was modeled using the TMM, which rigorously solves Maxwell’s equations for stratified media. Each layer $i$ with refractive index $n_i$, thickness $d_i$, and propagation constant $k_{zi}$ is represented by
\begin{equation}
M_i =
\begin{pmatrix}
\cos(k_{zi} d_i) & \frac{i}{q_i} \sin(k_{zi} d_i) \\
i q_i \sin(k_{zi} d_i) & \cos(k_{zi} d_i)
\end{pmatrix},
\end{equation}
where $q_i = n_i \cos\theta_i$ for s-polarization and $q_i = \cos\theta_i / n_i$ for p-polarization. The total transfer matrix $M = \prod_i M_i$ provides the reflection and transmission coefficients. The computed spectra reproduce the experimentally observed resonance shifts when VO$_2$ optical constants for each phase are used.

The dielectric stack comprising the DBR consisted of alternating high- and low-index materials (e.g., $\mathrm{Ta_2O_5}$ and $\mathrm{SiO_2}$), which was deposited on top of a  63-nm VO$_2$ film on a thick quartz substrate. The design was optimized for a stop band centered around 800~nm. Each layer thickness was chosen to satisfy a quarter-wave optical condition. The DBR was constructed using a 4+1 unit cell design. Four symmetric unit cells of the form 
$[d_A/2, d_B, d_A/2]$ were used, followed by a defect cell $[d_A/2, d_B, d_C]$, where
\begin{equation}
    d_i = \frac{\lambda_c}{4n_i}
\end{equation}
and $i=A,B, \text{or } C$, and $d_C=\gamma d_A$ with $\gamma=1.57$ represents a scaling parameter perturbing the standard half-wave defect thickness. The VO$_2$ layer was placed directly adjacent to the defect region to maximize field overlap with the Tamm mode.


VO$_2$ thin films were purchased from YANNA New Material Technology Co. Ltd., China, and grown to 63 nm on quartz substrates. Multilayers comprising the DBR consisting of Ta$_2$O$_5$/SiO$_2$ pair were prepared on the VO$_2$ films with the corresponding structures using electron beam evaporation (AdNaNotek EBS-150U);

The complex optical constants ($\tilde{n} = n + ik$) of the VO$_2$ thin films were characterized using a spectroscopic ellipsometer over a spectral range of $\lambda=[400,2500]$ nm, shown in Fig. S1. To capture the optical evolution through the IMT, samples were mounted on a thermal controller (TEC) and measured at discrete temperatures $T=[20,30,40,50,55,60,65,70,80,90]$. A stabilization period of one minute was maintained at each setpoint to ensure thermal equilibrium across the film before data acquisition. The ellipsometric parameters, $\Psi$ and $\Delta$, were analyzed using a multi-layer model. The dielectric function of the VO$_2$ layer was parameterized using Cauchy-Lorentz oscillators for the insulating state and a Drude-Lorentz model for the metallic state. The final $n$ and $k$ spectra were extracted by minimizing the Mean Squared Error (MSE) between the experimental data and the physical model.

\begin{figure}[t!]
    \centering
\includegraphics[width=0.43\textwidth]{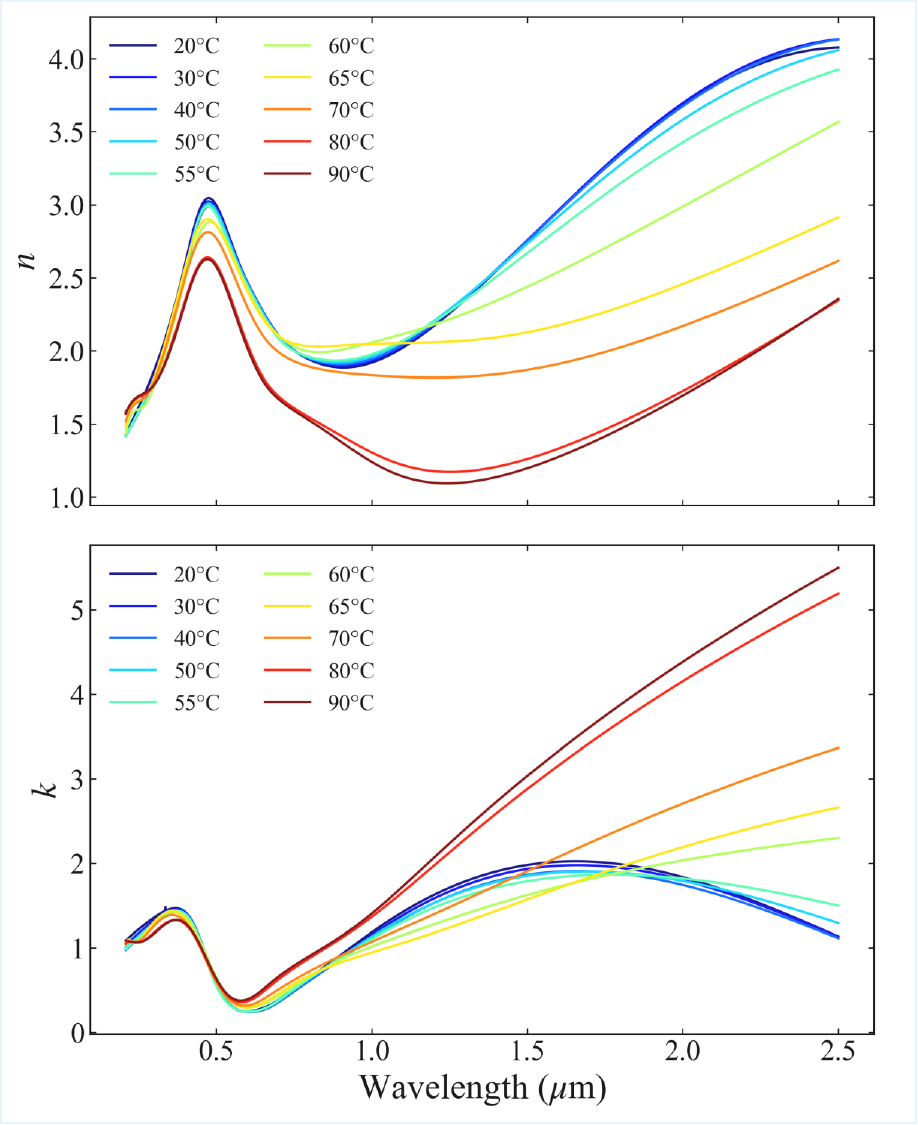}
\caption{\textbf{Ellipsometric measurements of VO$_2$ thin films during the phase transition.} Measured ellipsometric parameters $\Psi$ (amplitude ratio) and $\Delta$ (phase difference) are shown across a broad spectral range from 0.4 to 2.5 $\mu$m. Data were collected at ten discrete temperatures ranging from $20^\circ$C to $90^\circ$C to capture the evolution of the material's optical response during the insulator-to-metal transition (IMT). These measurements serve as the basis for extracting the temperature-dependent complex refractive index ($n + ik$) used in the numerical modeling of the OTS resonance.}
    \label{fig:FigureS1}
\end{figure}


The measurements were carried out at a constant temperature using an R1 angle-resolved spectrum system combined with a broadband white-light source and a PG4000 high-resolution spectrometer (Ideaoptics, China). The VO$_2$-DBR samples were placed on a thin electrically heated ceramic heating pad for heating. Temperatures were monitored by thermocouple filaments and controlled by a PID thermostat to drive the VO$_2$ transitions with an accuracy of $\pm 0.1^\circ$C. The evolution of the Tamm resonance was recorded across repeated heating and cooling cycles to assess reversibility. Experimental spectra were compared with TMM calculations to confirm the resonance position, linewidth, and field localization in each phase. The corresponding 2-d results for each temperature are shown in Fig. S2 to indicate the DBR and better highlight the movement of the OTS resonance. Note that this is equivalent to Fig. \ref{fig:Figure4}, but better highlights each spectral trace, as well as indicates the resonances are bound in the stopband.


The OTS dispersion reveals the profound impact of the VO$_2$ phase transition on the boundary conditions of the metastack. As shown in Fig. \ref{fig:Figure7}, the OTS exhibits a characteristic quadratic dependence on the incidence angle. This parabolic dispersion arises from the in-plane momentum of the mode, where the resonance condition blueshifts as the transverse wavevector increases.

\begin{figure*}
    \centering
\includegraphics[width=0.8\textwidth]{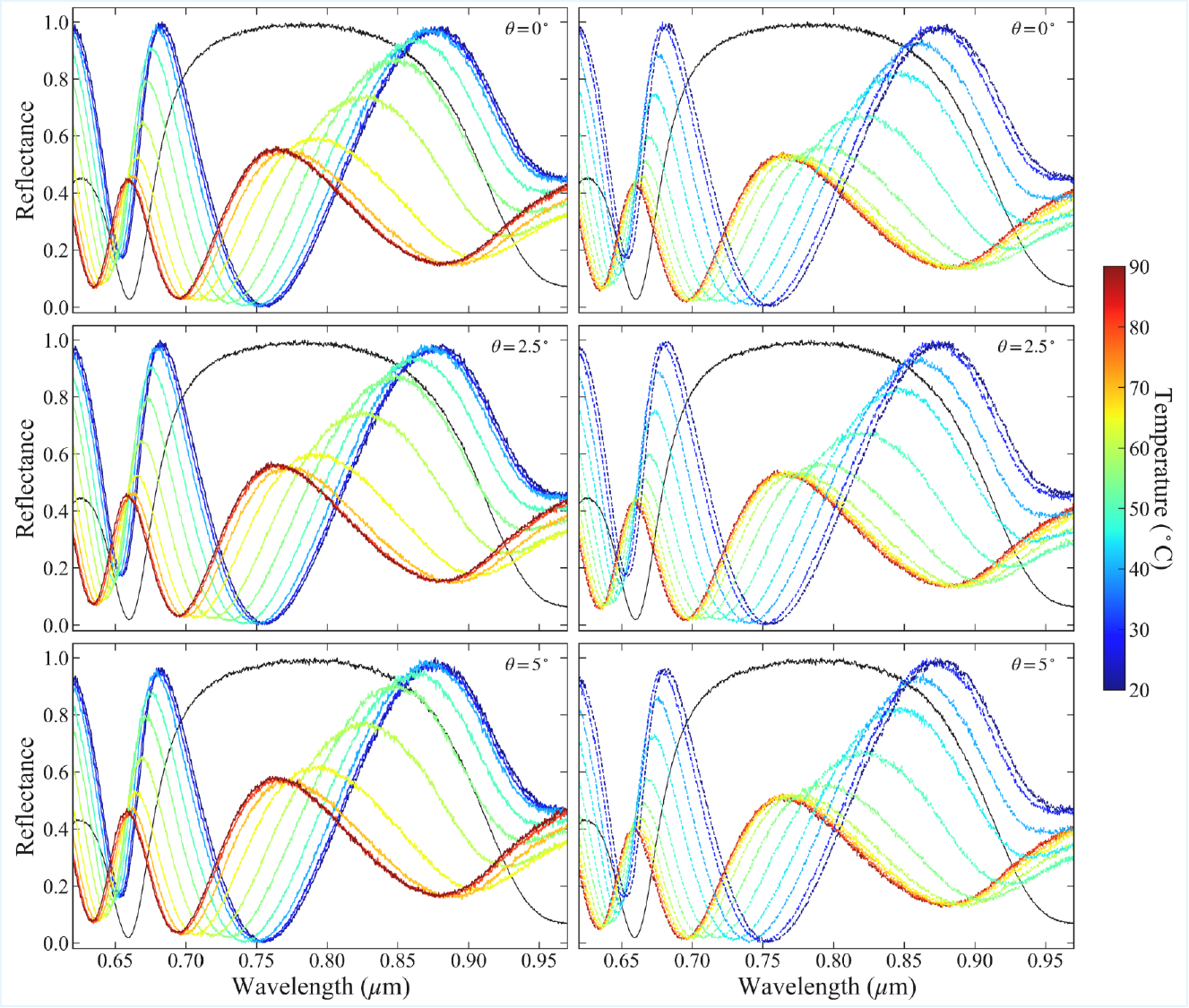}
\caption{\textbf{Reflectance spectra of the DBR and the VO$_2$ metastack} Two-dimensional slices for each independent temperature of the reflectance data (as presented in Fig. \ref{fig:Figure4}) are provided for heating (left column) and cooling (right column) cycles. These plots highlight the static position of the DBR stopband—represented by the black reference curves against the dynamic shift of the OTS resonance. The spectra are shown for incidence angles of $\theta = 0^\circ$, $2.5^\circ$, and $5^\circ$ to demonstrate the angular dependence of the resonance, as well as the movement of the resonance across IMT over temperature range of $20^\circ$C to $90^\circ$C.}
    \label{fig:FigureS2}
\end{figure*}

\begin{figure*}
    \centering   
    \includegraphics[width=0.85\textwidth]{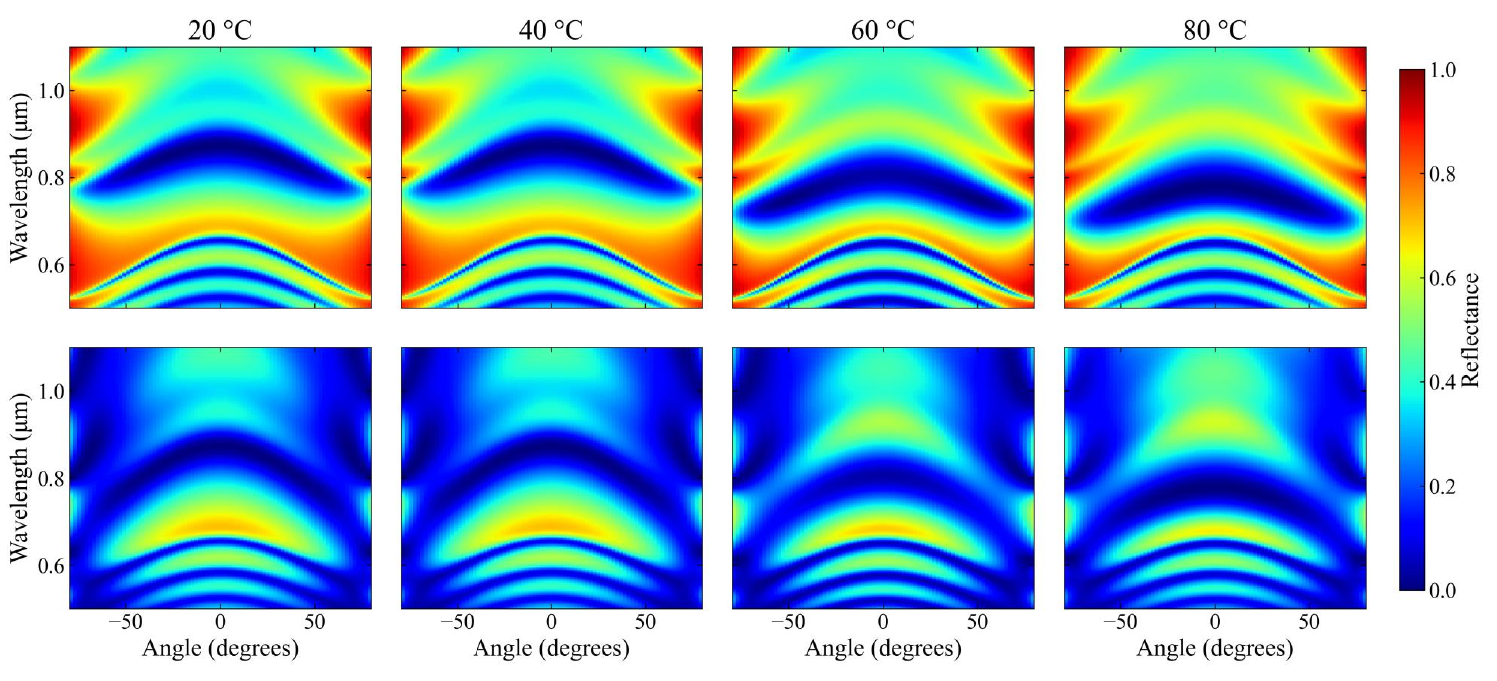}
    \caption{\textbf{VO$_2$-OTS Dispersion Response for s- and p-Polarizations.}  OTS dispersion is shown for four temperatures,  $T=[20, 40, 60, 80]^\circ C$, spanning wavelengths from 0.5 to 1.1 $\mu$m and angles from 0 to 80 degrees for s- (p-) polarization on the top (bottom) row. A quadratic dependence on the dispersion is observed for the OTS.}
    \label{fig:Figure7}
\end{figure*}

\end{document}